\documentclass[12pt]{iopart}
\usepackage{iopams}  
\usepackage{graphicx,bm}

\def\dd{\mathrm{d}}
\def\ii{\mathrm{i}}
\def\ee{\mathrm{e}}

\begin{document}

\title[Synthetic topological Kondo insulator in a pumped optical cavity]{Synthetic topological Kondo insulator in a pumped optical cavity}
\author{Zhen Zheng, Xu-Bo Zou and Guang-Can Guo}
\address{Key Laboratory of Quantum Information, and Synergetic Innovation Center of Quantum Information and Quantum Physics, University of Science and Technology of China, Hefei, Anhui, 230026, People's Republic of China}
\ead{xbz@ustc.edu.cn}
\vspace{10pt}

\begin{abstract}

Motivated by experimental advances on ultracold atoms coupled to a pumped optical cavity,
we propose a scheme for synthesizing and observing the Kondo insulator in Fermi gases trapped in optical lattices.
The synthetic Kondo phase arises from the screening of localized atoms coupled to mobile ones,
which in our proposal is generated via the pumping laser as well as the cavity.
By designing the atom-cavity coupling,
it can engineer a nearest-neighbor-site Kondo coupling that plays an essential role for supporting topological Kondo phase.
Therefore, the cavity-induced Kondo transition is associated with a nontrivial topological features,
resulting in the coexistence of the superradiant and topological Kondo state.
Our proposal can be realized with current technique, and thus
has potential applications in quantum simulation of the topological Kondo insulator in ultracold atoms.

\end{abstract}

%
\vspace{2pc}

\section{Introduction}

The experimental realization and manipulation of ultracold atoms
provide a versatile platform with feasible controllability to simulate the many-body physics in strongly correlated systems \cite{quantum-simulation}.
Taking advantages of current technique,
theoretic and experimental investigations on ultracold atoms advance in creating artificial gauge fields 
\cite{gauge-field-1,gauge-field-2,gauge-field-3,gauge-field-4,gauge-field-5}
and a variety of optical lattices \cite{sp-lattice,exotic-lattice-1,exotic-lattice-2,exotic-lattice-3},
which opens the way to explore and predict unconventional properties of
strongly correlated systems, such as the topological the Kondo insulator \cite{topo-kondo-rev-1,topo-kondo-rev-2}.

The Kondo phase arises from the screening of localized electrons hybridized with mobile ones,
forming a so-called Kondo insulator \cite{1d-kondo-rev}.
It captures many unusual properties of strongly correlated systems such as the heavy fermion materials.
With coexistence of the spin-orbital coupling, it brings rich physics with the discovery of the topological Kondo insulator
\cite{topo-kondo-1,topo-kondo-2,topo-kondo-3,topo-kondo-4}.
Recent studies of ultracold atoms shed light on the realization of the Kondo insulator
by proposing a variety of schemes that are widely based on
the optical coupling \cite{laser-kondo},
the Bloch-orbital hybridizations \cite{topo-kondo-5}, and the orbital Feshbach resonance 
\cite{orbit-kondo-1,orbit-kondo-2,soc-kondo,orbit-kondo-3,orbit-kondo-4,orbit-kondo-5}.
Especially in the earlier work \cite{topo-kondo-5},
artificial gauge fields generated by the laser-induced Raman coupling
make it possible to detect the topological Kondo insulator.
On the other hand, the study of Fermi gases in an optical cavity \cite{cavity-rev,cavity-1,cavity-2,cavity-3,cavity-4,cavity-5}
offers another framework for engineering artificial gauge fields 
\cite{soc-cavity-1,soc-cavity-2,magnetic-cavity-1,magnetic-cavity-2,magnetic-cavity-3,magnetic-cavity-4}.
It predicts the existence of the topological superradiant states \cite{topo-sr-1,topo-sr-2,topo-sr-3}.
This motivates us to find an alternative scheme to synthesize a topological Kondo insulator 
by implementing the cavity field.

Here in this paper, we present a proposal for realizing the synthetic topological Kondo insulator in a pumped optical cavity.
The paper is organized as follows.
In Section \ref{sec-model},
we start with the Hamiltonian describing atoms coupled to a pumped optical cavity,
and design it to obtain the Kondo lattice Hamiltonian.
In Section \ref{sec-result},
we show the phase transition by changing experimental controllable parameters,
and show the topological features of the superradiant Kondo phase.
The extension of our proposal to a higher-dimensional case are discussed in Section \ref{sec-discuss}.
In Section \ref{sec-con}, we summarize the work.
The details of the effective Hamiltonian and the slave boson approach, which we use to study the Kondo phase, 
are formulated in \ref{sec-app-h} and \ref{sec-app-sb}.

\section{The model}\label{sec-model}
\subsection{Effective Hamiltonian} \label{sec-hamiltonian}

\begin{figure}[tbp]
\centering
\includegraphics[width=0.8\textwidth]{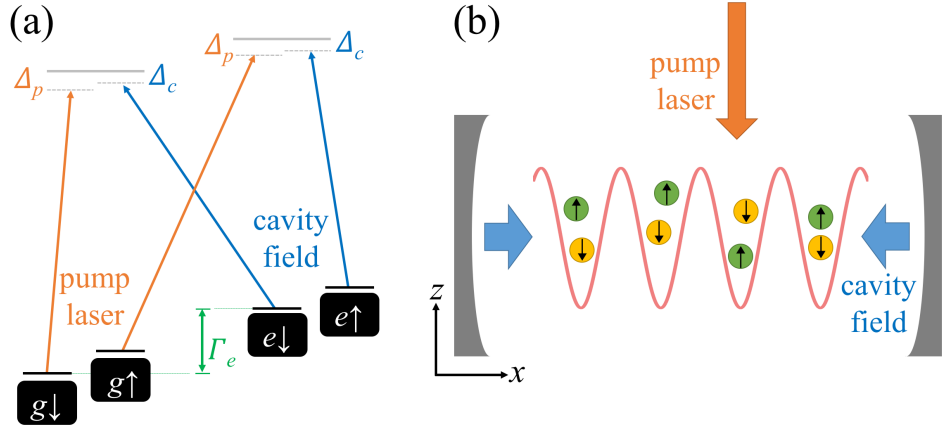}
\caption{(a) Illustration of the atomic transition.
(b) Illustration of the experimental proposal.
The fermionic atoms confined in the optical lattice
are coupled to the cavity oriented in $x$ axis in accompany with a pumping laser in the $x$-$z$ plane.}
\label{fig-setup}
\end{figure}

We consider ultracold fermionic atoms trapped in a one-dimensional (1D) optical lattice oriented in $x$ direction.
The atomic level structure is illustrated in Figure. \ref{fig-setup}(a).
In practice, for $^6$Li alkali atoms as an example, 
we can choose two nuclear spin states with $|F,m_F\rangle=|1/2,1/2\rangle$ and $|1/2,=-1/2\rangle$ as $|g,\uparrow\downarrow\rangle$,
and $|3/2,3/2\rangle$ and $|3/2,1/2\rangle$ as $|e,\uparrow\downarrow\rangle$, respectively.
The spinful atoms are initially prepared in $|g\rangle$ with double filling in each site.
The experimental setup for our proposal is sketched in Figure. \ref{fig-setup}(b).
When placed into a high-finesse cavity,
the Raman transition between $|g,\sigma\rangle$ and $|e,\sigma\rangle$ ($\sigma=\uparrow,\downarrow$)
is driven by a plane-wave pumping laser $h(\bm{r})$ with frequency $\omega_p$ and linear polarization in accompany with
a single-mode standing-wave cavity field $\eta(x)$ with frequency $\omega_c$ and $\sigma^-$ polarization.
The selection rule in the Raman transition can suppress the unwanted atomic transitions.
The optical cavity is oriented in the $x$ axis,
while the pumping laser is placed in the $z$ plane.
In order to realize the Raman transition by using the same laser and cavity fields,
we introduce an AC-Stark shift individually to $|g,\downarrow\rangle$, which can be generated by a far-detuned laser.
This system is described by a Hamiltonian composed of three terms,
\begin{equation}
H = H_A+H_C+H_I ~. \label{eq-hamilton0}
\end{equation}
The first term $H_A$ describes the atom subsystem,
\begin{equation}
H_A = \int \dd x \,\Big\{ H_{L}(x) - \sum_{\sigma}\Delta_{a\sigma} \psi_{g\sigma}^\dag(x)\psi_{g\sigma}(x) \Big\} ~.
\end{equation}
Here
\begin{equation}
H_{L}(x)=\sum_{\lambda,\sigma}\psi_{\lambda\sigma}^\dag(x) \Big[ -\frac{\nabla^2}{2m} 
+ V_\lambda(x) \Big] \psi_{\lambda\sigma}(x)
\end{equation}
describes free atoms confined in the lattice.
$\{\psi_{\lambda\sigma},\psi_{\lambda\sigma}^\dag\}$
are the annihilation and creation field operators for the level $\lambda=g/e$, respectively.
$V_\lambda(x)=V_\lambda\sin^2(k_Lx)$ is the optical lattice trap potential where $k_L=\pi/d$ with $d$ as the lattice constant.
$\Delta_{a\sigma}$ is the detuning between $|g,\sigma\rangle$ and $|e,\sigma\rangle$.
For simplicity without loss of generality, we assume $\Delta_{a\uparrow}=\Delta_{a\downarrow}=\Delta_{a}$.
In the whole paper we set $\hbar=1$.
The second term $H_C$ describes the cavity subsystem,
\begin{equation}
H_C = -\Delta_c a^\dag a ~.
\end{equation}
Here $\{a,a^\dag\}$ are the annihilation and creation operators for cavity fields, respectively.
$\Delta_c$ is the cavity-pump detuning.
The last term $H_I$ describes the interaction between the atom and cavity subsystems,
\begin{equation}
H_I = \int \dd x \, g(x) \sum_{\sigma}[ a \psi_{e\sigma}^\dag(x) \psi_{g\sigma}(x) + H.c.] ~.
\end{equation}
Here $g(x)$ denotes the atom-cavity coupling mode, which is originated from $h(\bm{r})$ and $\eta(x)$.
The term $H.c.$ stands for Hermitian conjugation.
The details of the Hamiltonian (\ref{eq-hamilton0}) are given in \ref{sec-app-h}.

From the Hamiltonian (\ref{eq-hamilton0}), taken the cavity decay into considerations,
the Heisenberg equation for the cavity fields $a$ is written as
\begin{equation}
\ii\partial_t a = [a,H]=
- (\Delta_c+\ii\kappa) a + g(x) \sum_{\sigma}\psi_{g\sigma}^\dag(x)\psi_{e\sigma}(x) ~, \label{eq-steady}
\end{equation}
where $\kappa$ is the decay rate.
Hereafter we make the mean-field approximation $\langle a \rangle=\alpha$.
As known in previous works \cite{cavity-5},
the system undergoes a superradiant transition charactered by $\alpha$,
which is driven by tuning experimental parameters, \textit{e.g.} the pumping laser amplitude $\Omega_p$.
When $\alpha$ is nonzero, the system is noted as the superradiant state.
Due to the cavity decay, the system is indeed a non-equilibrium one.
For steady cavity field, $\alpha$ satisfies $\partial_t \alpha = 0$, and thus we obtain
\begin{equation}
\alpha = \frac{\eta}{\Delta_c+\ii\kappa} ~,\qquad
\eta = \frac{1}{L}\int \dd x \sum_{\sigma} g(x) \langle\psi_{g\sigma}^\dag(x)\psi_{e\sigma}(x)\rangle ~. \label{eq-super-rad}
\end{equation}
Here $L$ is the length of the 1D lattice.
Inserting $\alpha$ back into Eq.(\ref{eq-hamilton0}), the Hamiltonian is recast as
\begin{equation}
\hat{H} = \int \dd x \,\sum_{\sigma} \Big\{ H_L(x)
-\Delta_{a} \psi_{g\sigma}^\dag(x) \psi_{g\sigma}(x)
+\big[ \alpha g(x) \psi_{e\sigma}^\dag(x) \psi_{g\sigma}(x) + H.c.\big] \Big\} ~. \label{eq-harmilton_hat}
\end{equation}

In order to derive a Hubbard Hamiltonian to quantum simulate the Kondo physics,
we use the tight-binding approximation and
expand $\psi_{\lambda\sigma}(x)$ in terms of Wannier wave functions $W_\lambda(x)$.
Besides, the following two manipulations are required:
(i) atoms in $|g\rangle$ are deeply trapped in the lattice to make the tunneling amplitude $t_g\ll t_e$.
This can be achieved by means of two groups of counter-propagate lasers to 
create optical lattices separately trapping $|g\rangle$ and $|e\rangle$ with different depth.
(ii) Via Feshbach resonance, we introduce strong repulsive interaction between $|g\rangle$ with opposite spins in the same site.
For example of $^6$Li \cite{feshbach-rev}, we can make broad Feshbach resonance
between $|g,\uparrow\rangle$ and $|g,\downarrow\rangle$.
By contrast in the same parameter region, the scattering between $|e,\uparrow\rangle$ and $|e,\downarrow\rangle$ is off resonance,
hence their interaction is ignorable.
After accomplishing the above manipulations,
$|g\rangle$ can simulate the localized fermions in Kondo problems,
while $|e\rangle$ simulates the conduction fermions.
The effective Hamiltonian is then expressed as
\begin{eqnarray}
\mathcal{H} =& -\sum_{\langle ij \rangle}\sum_{\lambda,\sigma} t_\lambda \lambda_{i\sigma}^\dag \lambda_{j\sigma}
-\Delta_{a} \sum_{j,\sigma} g_{j\sigma}^\dag g_{j\sigma}
+ U\sum_j g_{j\uparrow}^\dag g_{j\downarrow}^\dag g_{j\downarrow}g_{j\uparrow} \nonumber\\
&+ \sum_{ij,\sigma} (V_{ij} e_{i\sigma}^\dag g_{j\sigma} + H.c. ) ~, \label{eq-hamitonian-tran-before}
\end{eqnarray}
where $\sum_{\langle ij \rangle}$ denotes the summation between nearest-neighbor sites,
$\{ \lambda_{j\sigma}, \lambda_{j\sigma}^\dag \}$ are the annihilation and creation operators for atoms on $j$-th site,
and $U$ is the repulsive interaction strength.
The synthetic Kondo coupling $V_{ij}\equiv \alpha J_{ij}$ with
$J_{ij} = \int \dd x\, g(x)W_e^*(x-x_i) W_g(x-x_j)$.

From \cite{topo-kondo-rev-2}, we know that, aiming to synthesize the gapped Kondo state, \textit{i.e.} the Kondo insulator,
it is required that the tunneling of the localized fermions hosts an opposite sign compared with the conduction fermions.
Unfortunately, this requirement is not easily satisfied,
because the signs of the tunneling for the lowest Bloch bands, \textit{i.e.} $t_g$ and $t_e$, are usually the same.
On the other hand, if we invoke the following transformation
\begin{equation}
\tilde{g}_{j\sigma} = g_{j\sigma} \ee^{\ii j\pi} \label{eq-tran}
\end{equation}
into the Hamiltonian (\ref{eq-hamitonian-tran-before}),
the atom fields $\tilde{g}$ and $e$ will host tunneling with opposite signs.
However, the new operator representation may lead to a staggered Kondo coupling $(V_{ij} \ee^{\ii j\pi}e_{i\sigma}^\dag \tilde{g}_{j\sigma}+H.c.)$.
Next, we design the pumping laser and cavity mode to eliminate the staggered phase.

\subsection{Synthetic Kondo coupling} \label{sec-couple}

The synthetic Kondo coupling $V_{ij}=\alpha K_{ij}$ is generated by the pumping laser as well as the atom-cavity coupling.
It is obviously proportional to the superrandiance order $\alpha$,
hence works only in the superrandiant state.
As we focus on the physics of the 1D lattice system along the $x$ axis,
the plane-wave mode of the pumping laser can be approximately treat as a constant strength $h(\bm{r})=\Omega_p$.
From \cite{topo-kondo-rev-2}, we know the on-site Kondo coupling does not introduce topological nontrivial properties to the system.
On the other hand, the beyond-on-site components with odd parity will exhibit features of spin-orbital couplings,
and host the possibility to harbor a topological Kondo state.
For this sake, we design the standing-wave mode of the cavity as
$\Omega(x)=\Omega_c \sin(k_c x)$ where $\Omega_c$ is the atom-cavity coupling strength and
$k_c$ is the cavity mode momentum.
In this way, the on-site component in $J_{ij}$ vanishes while 
the nearest-neighbor-site component dominates.
When we tune $k_c$ to match $k_L/2$, it will impose a phase 
$\ee^{-\ii j\pi}$ into $J_{ij}$, 
eliminating the staggered phase $\ee^{\ii j\pi}$ generated by the new operator representation (\ref{eq-tran}).
Thus we can obtain $J_{ij}\approx \pm K\delta_{i,j\pm1}\ee^{-\ii j\pi}$.
Here we denote $K\equiv \tilde{\Omega}\int \dd x\, \sin(k_Lx/2)W_e^*(x) W_g(x-a)$.
$\tilde{\Omega}$ is the coupling strength between $|g\rangle$ and $|e\rangle$,
and its detailed form is given in \ref{sec-app-h}.

Based on the above designs,
the final expression of the effective Hamiltonian is written as
 \begin{eqnarray}
\mathcal{H} =& \sum_{\langle ij \rangle,\sigma} \big( t_g\tilde{g}_{i\sigma}^\dag \tilde{g}_{j\sigma} - t_e e_{i\sigma}^\dag e_{j\sigma} \big)
-\Delta_{a} \sum_{j,\sigma} \tilde{g}_{j\sigma}^\dag \tilde{g}_{j\sigma}
+ U\sum_j \tilde{g}_{j\uparrow}^\dag \tilde{g}_{j\downarrow}^\dag \tilde{g}_{j\downarrow} \tilde{g}_{j\uparrow} \nonumber\\
&+ \sum_{\langle ij \rangle,\sigma} (V_{ij} e_{i\sigma}^\dag \tilde{g}_{j\sigma} + H.c. ) ~, \label{eq-hamitonian-final}
\end{eqnarray}
where $V_{ij}=\pm \alpha K\delta_{i,j\pm1}$.
At the limit $t_g\ll t_e$ and $U\rightarrow \infty$, the Hamiltonian (\ref{eq-hamitonian-final}) 
describes the Kondo lattice model associated with the off-site Kondo coupling \cite{topo-kondo-rev-2}.

\subsection{Slave boson method}\label{sec-method}

We employ the slave boson method \cite{1d-kondo-rev} 
to explore the possible Kondo phase arising from our model Hamiltonian (\ref{eq-hamitonian-final}).
In the slave boson method, the localized fermion operator $\tilde{g}$ are decomposed into
a fermion operator $\hat{g}$ associated with an auxiliary boson operator $\hat{b}$,
\begin{equation}
\tilde{g}_{j\sigma}=\hat{g}_{j\sigma} \hat{b}_j^\dag ~.
\end{equation}
In this paper, we make the mean-field approximation to the boson operator 
$\hat{b}_j=\langle \hat{b}_j \rangle \equiv b_j \approx b$ and
$\hat{b}_j^\dag=\langle \hat{b}_j ^\dag \rangle \equiv b_j^* \approx b^*$,
and recognize $b$ as the order parameter of the emergent Kondo phase.
We focus on the results at zero-temperature limit,
because the mean-field method can capture the qualitative and the topological features of the lattice system.
After processing the standard approach (see \ref{sec-app-sb}), we can get the effective action
\begin{equation}
S= \int_0^{\beta} \dd \tau \,\Big[ \sum_{j\sigma} \big( 
\hat{g}_{j\sigma}^\dag \partial_\tau \hat{g}_{j\sigma} 
+ e_{j\sigma}^\dag \partial_\tau e_{j\sigma} \big)
- \mathcal{H}_\mathrm{eff} \Big] ~.
\end{equation}
Here $\tau\equiv\ii t$ is the imaginary time.
$\beta\equiv1/(k_BT)$ with temperature $T$ and we set the Boltzmann constant $k_B=1$ in the whole paper.
$\mathcal{H}_\mathrm{eff}$ is expressed as \cite{kondo-book}
\begin{eqnarray}
\mathcal{H}_\mathrm{eff} =& E_0+t_g\sum_{\langle ij \rangle,\sigma} \hat{g}_{i\sigma}^\dag \hat{g}_{j\sigma}
-t_e\sum_{\langle ij \rangle,\sigma} e_{i\sigma}^\dag e_{j\sigma}
+\sum_{j,\sigma}(\lambda_j-\Delta_{a})\hat{g}_{j\sigma}^\dag \hat{g}_{j\sigma} \nonumber\\
&+\sum_{\langle ij \rangle,\sigma} (V_{ij} e_{i\sigma}^\dag b_j^*\hat{g}_{j\sigma} + H.c. ) \label{eq-hamit-eff}
\end{eqnarray}
with $E_0\equiv \sum_{j} \lambda_j ( |b_{j}|^2 -1)+\Delta_c |\alpha|^2$.
$\lambda_j$ is the Lagrange multiplier introduced by the number conservation on each site,
and we can see $\Delta_{a}-\lambda_j$ characterizes the effective chemistry potential of the localized $\hat{g}$ fermions.
The ground state at zero temperature limit can be determined by self-consistently minimizing the free energy $F$
with respect to the order parameter $b$ and $\lambda$ (See \ref{sec-app-sb}):
(i) When $b$ is nonzero, the system is a Kondo state,
and moreover is a superradiant Kondo state if $\alpha\neq0$.
(ii) When $b$ vanishes, the system is a normal gas state.

\section{Phase transition and topological features}\label{sec-result}

\begin{figure}[tbp]
\centering
\includegraphics[width=0.7\textwidth]{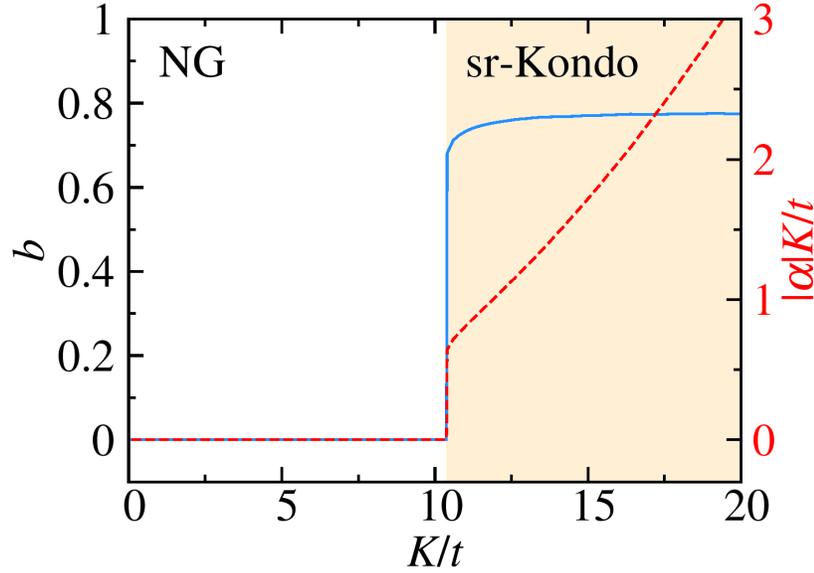}
\caption{The Kondo order $b$ (blue solid line) and the cavity-induced coupling $\alpha K$ (red dashed line) as functions of $K$.
NG stands for the normal gas state and sr-Kondo for the superradiant Kondo state, respectively.
For convenience, we use the tunneling amplitude $t\equiv t_e$. 
Other parameters are $t_g=0.2t$, $\Delta_a=0.35t$, $\Delta_c=-5t$, and $\kappa=0.01t$.}
\label{fig-phase}
\end{figure}

In Figure. \ref{fig-phase}, we plot the Kondo order $b$ and the superradiance order $\alpha$ with respect to $K$ at zero temperature limit.
Experimentally, $K$ can be tuned by the pumping laser strength $\Omega_p$ and the configuration of the optical cavity.
The gap of the Kondo phase is the manifest signature distinct from the normal gas state,
resulting in a Kondo insulator \cite{1d-kondo-rev}.
It originates from the screening of localized $\hat{g}$ fermions coupled to the mobile $e$ fermions,
and is evaluated by the nonzero order parameter $b$.
In Figure. \ref{fig-phase} we can see there exist two states as the ground state of the lattice system.
When $K$ exceeds a threshold,
the system simultaneously processes two types of phase transitions: the Kondo transition from the normal gas state to the gapped Kondo phase,
and the superradiant transition from the normal gas state to the superradiant state.
This is because in Section \ref{sec-couple} we have know the Kondo coupling works only in a superradiant state.
In another words, the atom-cavity coupling is responsible to the Kondo screening.
As both the two order parameters exhibit discontinuous evolutions across the phase transition,
it yields that both the Kondo and superradiant transitions are of first order.

What interests us more here is whether the cavity-induced superradiant Kondo phase is a topological nontrivial state.
To analyze its possible topological properties,
we make a Fourier transformation to the effective Hamiltonian (\ref{eq-hamit-eff}).
In the base $\Psi_k= (\hat{g}_{k\uparrow},\hat{g}_{k\downarrow},e_{\uparrow},e_{\downarrow})^T$, it is written as
\begin{equation}
\mathcal{H}_\mathrm{eff}(k)=\left(\begin{array}{cccc}
\epsilon_g(k)  & bV_k^* \\
b^*V_k & \epsilon_e(k)
\end{array}\right)\otimes\mathrm{I}_{2\times2} ~, \label{eq-hamit-eff-k}
\end{equation}
where $\epsilon_g(k) = 2t_g \cos(kd)+ \lambda - \Delta_a$,
$\epsilon_e(k) = -2t_e\cos(kd)$,
$V_k= -\ii 2 \alpha K\sin(kd)$,
and $\mathrm{I}_{2\times2}$ is the 2$\times$2 identity matrix.
We treat $b$ as a real number here,
since its phase can be rotated off in $\mathcal{H}_\mathrm{eff}$ and does not change the physics.
The Hamiltonian (\ref{eq-hamit-eff-k}) can be decompose into two parts:
$\mathcal{H}_\mathrm{eff}=\mathcal{H}_0+\mathcal{H}_1$ with the notations
$\mathcal{H}_0(k) \equiv \frac{\epsilon_g(k)+\epsilon_e(k)}{2}\mathrm{I}_{4\times4}$
and $\mathcal{H}_1(k) \equiv (\bm{d}_k\cdot \bm{\sigma})\otimes\mathrm{I}_{2\times2}$.
Here $\bm{d}_k=(d_k^x,d_k^y,d_k^z)$ with
$d_k^x=0$, $d_k^y=-2b\alpha K\sin(kd)$, and
$d_k^z = \frac{\epsilon_g(k)-\epsilon_e(k)}{2}$.
$\sigma_i$ ($i=x,y,z$) are Pauli matrices.
By making a unitary transformation $\mathcal{U}=\exp(\ii \mathcal{H}_0 t)$,
we can eliminates $\mathcal{H}_0$ in $\mathcal{H}_\mathrm{eff}$, and obtain
\begin{equation}
\widetilde{\mathcal{H}}_\mathrm{eff} 
= \mathcal{U}\mathcal{H}_\mathrm{eff}\mathcal{U}^\dag - \mathcal{U}\ii \partial_t\mathcal{U}^\dag
= \mathcal{H}_1 + [\mathcal{H}_0, \mathcal{H}_1] + \frac{1}{2}[\mathcal{H}_0, [\mathcal{H}_0, \mathcal{H}_1] ] + \cdots
\end{equation}
where we have used the Campbell-Baker-Hausdorff expansion.
It is easy to demonstrate $[\mathcal{H}_0, \mathcal{H}_1]=0$ since $\mathcal{H}_0$ behaves like an identity matrix,
and then we obtain $\widetilde{\mathcal{H}}_\mathrm{eff} = \mathcal{H}_1$.
Therefore, the lattice system described by $\mathcal{H}_\mathrm{eff}$ 
shares the same topological properties from the eigenstates of $\mathcal{H}_1$ \cite{topo,topo-rm}.
We can see that the Hamiltonian $\mathcal{H}_1$ respects the particle-hole symmetry:
$\Xi \mathcal{H}_1(k) \Xi^\dag = -\mathcal{H}_1(-k)$, where $\Xi=\sigma_x\mathcal{K} \otimes \mathrm{I}_{2\times2}$ 
and $\mathcal{K}$ is the complex conjugate operator.
It indicates the superradiant Kondo phase belongs to a D topological class \cite{topo-kondo-cri},
which is characterized by a $\mathbb{Z}_2$ topological invariant \cite{topo-kondo-coleman}.

\begin{figure}[tbp]
\centering
\includegraphics[width=0.8\textwidth]{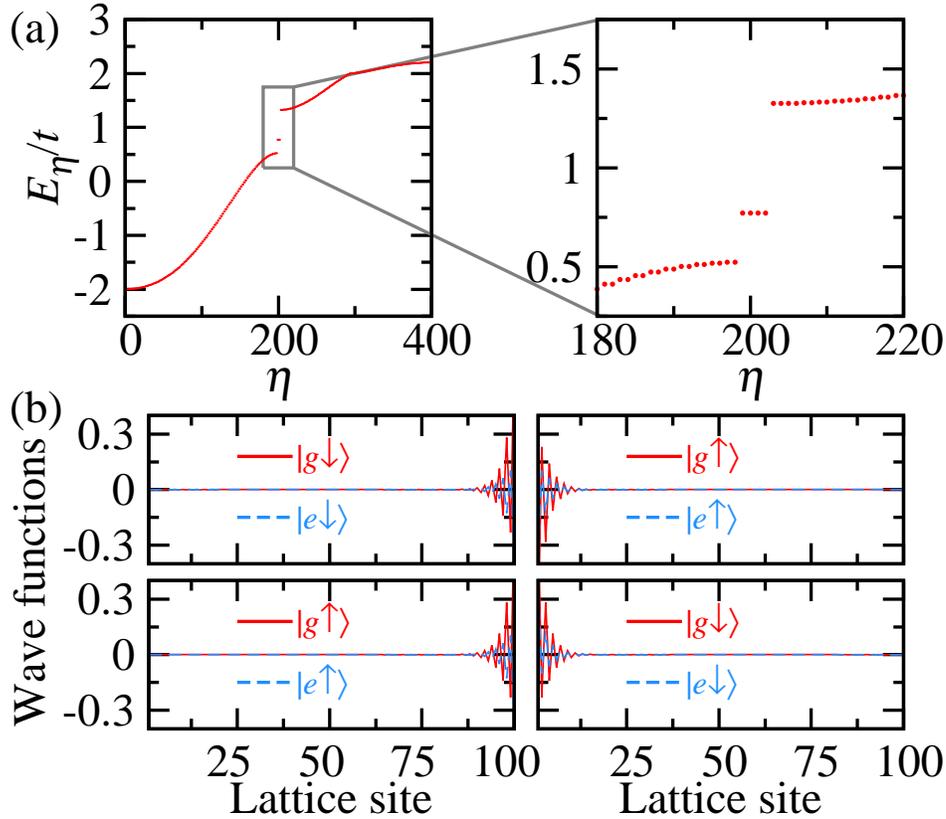}
\caption{(a) Energy spectrum of the superradiant Kondo phase.
We set the lattice length $L=100$ and $K=12t$.
Other parameters are $t_g=0.2t$, $\Delta_a=0.35t$, $\Delta_c=-5t$, and $\kappa=0.01t$.
(c) Spatial distributions of the four edge states in (a).}
\label{fig-edge}
\end{figure}

In Figure. \ref{fig-edge}(a), we display the Bogoliubov-de Gennes (BdG) quasiparticle spectrum (see \ref{sec-app-sb})
for the superradiant Kondo phase.
Distinct from the trivial Kondo phase, inside the band gap,
the superradiant Kondo phase hosts edge states whose wave functions are localized on the chain ends, as shown in Figure. \ref{fig-edge}(b).
This is the key signature as a topological insulator for the superradiant Kondo phase.
The lattice model has no spin hybridization and is spin degenerate.
Therefore, the edge states are four-fold degenerate.
They are protected by the particle-hole symmetry discussed above.

\section{Discussions}\label{sec-discuss}

Our proposal is readily extended to a two-dimensional (2D) case.
It can be realized by engineering a 2D optical lattice and two optical cavities in the $x$-$y$ plane.
We assume the cavity modes
$\eta_1(x)=\Omega_c\sin(k_Lx/2)$ and
$\eta_2(y)=\Omega_c\sin(k_Ly/2)$.
Then the Kondo coupling originated from the Raman transition can give rise to a Chern Kondo insulator state,
which has also been proposed via Bloch-orbital hybridizations in ultracold Fermi gases \cite{topo-kondo-5}.
It is noted that, in the 2D extension of our proposal, 
the emergency of the Chern Kondo insulator state is still ascribed to the cavity field.

\section{Conclusions}\label{sec-con}

In summary, we propose a scheme for synthesizing the Kondo insulator in
ultracold Fermi gases placed in a pumped optical cavity.
The synthetic Kondo coupling originates from the Raman transition,
which in our proposal is driven by the pumping laser as well as the cavity field.
Distinguished from the trivial Kondo phase,
the atom-cavity coupling gives rise to a synthetic nearest-neighbor-site Kondo coupling,
and is the key ingredient for supporting the topological superradiant Kondo phase with edge states gapped from the bulk.
Our proposal is simple and reliable based on current experimental technique,
and can provide a versatile platform for quantum simulating and studying 
the many-body physics and topological phases of the Kondo insulator.

\section*{Acknowledgements}

This work was supported by
Young Scientists Fund of the National Natural Science Foundation of China (Grant No. 11704367),
National Natural Science Foundation of China (Grants No. 11474271 and No. 11674305),
National Postdoctoral Program for Innovative Talents of China (Grant No. BX201600147),
and National Key R\&D Program (Grants No. 2016YFA0301300 and No. 2016YFA0301700).

\appendix
\section{Effective Hamiltonian} \label{sec-app-h}

For the model illustrated in Figure \ref{fig-setup},
we start with Jaynes-Cummings Hamiltonian \cite{soc-cavity-1}:
\begin{equation}
H(x) = H_{L}(x)+H_{1}(x)+H_{2}(x) ~.
\end{equation}
The first term describes the free atoms confined in the optical lattice,
\begin{equation}
H_{L}(x)=\sum_{\lambda=g/e,\sigma}\psi_{\lambda\sigma}^\dag(x) \Big[ -\frac{\nabla^2}{2m} 
+ V_\lambda(x) \Big] \psi_{\lambda\sigma}(x)~.
\end{equation}
The second term describes the transition between $|g\rangle$ and $|s\rangle$ 
($|s\rangle$ denote levels with the highest energy in the Raman transition, \textit{i.e.} the upper grey levels in Figure \ref{fig-setup}(a)),
\begin{equation}
H_{1}(x) = \sum_{\sigma}\Gamma_{s} \psi_{s\sigma}^\dag(x)\psi_{s\sigma}(x)+
\sum_{\sigma}\big[h(\bm{r}) \ee^{-\ii\omega_p t} \psi_{s\sigma}^\dag(x) \psi_{g\sigma}(x) + H.c.\big] ~.
\end{equation}
Here $\Gamma_s$ is the atomic frequency difference between $|g\rangle$ and $|s\rangle$.
The last term describes the transition between $|e\rangle$ and $|s\rangle$,
\begin{equation}
H_{2}(x) = \sum_{\sigma}\Gamma_{e} \psi_{e\sigma}^\dag(x)\psi_{e\sigma}(x)
+\omega_c a^\dag a + \eta(x) \sum_{\sigma}\big[ a \psi_{s\sigma}^\dag(x) \psi_{e\sigma}(x) + H.c.\big] ~.
\end{equation}
Here $\Gamma_e$ is the atomic frequency difference between $|g\rangle$ and $|e\rangle$.
As we choose $|g\rangle$ and $|e\rangle$ by hyperfine states with the same nuclear spin,
$\Gamma_s$ and $\Gamma_e$ can be tuned to be approximately independent of the pseudo-spins $\sigma$.
We make a unitary transformation
\begin{equation}
U= \exp\Big\{ \ii \sum_{\sigma}\big( \Gamma_{s}\psi_{e\sigma}^\dag \psi_{s\sigma} +
\Gamma_{e} \psi_{e\sigma}^\dag\psi_{e\sigma} \big)
+\omega_c a^\dag a \Big\}~.
\end{equation}
In the rotating frame, the Hamiltonian $H(x)$ is rewritten as
\begin{eqnarray}
H'(x) &= U^\dag H(x) U - U^\dag \ii\partial_t U \nonumber\\
&= H_L+ \sum_\sigma \big[ h(\bm{r}) \ee^{\ii \Gamma_p t} \psi_{s\sigma}^\dag \psi_{g\sigma} 
+ \eta(x) \ee^{\ii \Gamma_c t} a \psi_{s\sigma}^\dag \psi_{e\sigma} + H.c.\big] ~,
\end{eqnarray}
where $\Gamma_p=\Gamma_s-\omega_p$ and $\Gamma_c=\Gamma_s-\Gamma_e-\omega_c$.
Here we make the notations $\Delta\equiv(\Gamma_p+\Gamma_c)/2$ and $\delta\equiv(\Gamma_p-\Gamma_c)/2$,
and the make the unitary transformation
\begin{equation}
U'= \exp\Big\{ \ii \delta \Big( a^\dag a - \sum_{\sigma} \psi_{g\sigma}^\dag \psi_{g\sigma} \Big)t\Big\}~.
\end{equation}
The Hamiltonian $H'(x)$ is rewritten as
\begin{eqnarray}
H''(x) &= U'^\dag H'(x) U' - U'^\dag \ii\partial_t U' \nonumber\\
&= H_L+ \delta a^\dag a - \delta\sum_{\sigma} \psi_{g\sigma}^\dag \psi_{g\sigma}
+\sum_\sigma \big[ h(\bm{r}) \ee^{\ii \Delta t} \psi_{s\sigma}^\dag \psi_{g\sigma}  \nonumber\\
&\quad + \eta(x) \ee^{\ii \Delta t} a \psi_{s\sigma}^\dag \psi_{e\sigma} + H.c.\big] ~,
\end{eqnarray}
Adiabatically eliminating $|s\rangle$, we obtain the final form of the effective Hamiltonian in Section \ref{sec-hamiltonian}:
\begin{equation}
\mathcal{H}(x) = H_L+ \delta a^\dag a - \delta\sum_{\sigma} \psi_{g\sigma}^\dag \psi_{g\sigma}
+ g(x) \sum_{\sigma}[ a \psi_{e\sigma}^\dag(x) \psi_{g\sigma}(x) + H.c.] ~,
\end{equation}
where $g(x)=h(\bm{r})\eta(x)/\Delta=\tilde{\Omega}\sin(k_L2/x)$ with $\tilde{\Omega}\equiv\Omega_p\Omega_c/\Delta$.
In Section \ref{sec-hamiltonian}, we set $\Delta_c=-\delta$.
$\Delta_a$ can be changed by $\delta$ associated with an AC-Stark shift generated by a far-detuned auxiliary laser.

The parameters we choose in the phase diagram are accessible in practice.
They can be estimated as follows.
The numeric calculation via the maximally localized Wannier functions gives the tunneling
$t\equiv t_e\approx0.143E_R$ with $V_e=2E_R$, and $t_g\approx0.0308E_R\approx 0.215t$ with the trap depth $V_g=8E_R$ 
Here $E_R\equiv \hbar^2/2md^2$ is the lattice recoil energy.
When tuning $\Delta\approx300E_R$, $\Omega_p\approx 100E_R$ and $\Omega_c\approx600t$,
it gives $K\approx 12.03t$.
Therefore, the parameter regions displayed in Figure \ref{fig-phase} are available in real ultracold Fermi gases.

\section{Slave boson approach} \label{sec-app-sb}

In the slave boson method, the localized fermion operator $\tilde{g}$ is written as a decomposition of
a fermion operator $\hat{g}$ as well as an auxiliary boson operator $\hat{b}$,
\begin{equation}
\tilde{g}_{j\sigma}=\hat{g}_{j\sigma} \hat{b}_j^\dag ~.
\end{equation}
Here $\hat{g}$ and $\hat{b}$ satisfy the single occupancy constraint on each site,
\begin{equation}
\hat{b}_j^\dag \hat{b}_j + \sum_\sigma \hat{g}_{j\sigma}^\dag \hat{g}_{j\sigma} = 1 ~.\label{eq-app-constraint}
\end{equation}
Then by excluding the $U$-term,
the Kondo lattice Hamiltonian (\ref{eq-hamitonian-tran-before}) can now be recast as
\begin{eqnarray}
\tilde{\mathcal{H}} =&~ t_g\sum_{\langle ij \rangle,\sigma} \hat{g}_{i\sigma}^\dag \hat{g}_{j\sigma}
-t_e\sum_{\langle ij \rangle,\sigma} e_{i\sigma}^\dag e_{j\sigma}
-\Delta_{a}\sum_{j,\sigma}\hat{g}_{j\sigma}^\dag \hat{g}_{j\sigma} \nonumber\\
&+ \alpha K\sum_{j,\sigma}\big( \hat{b}_j^\dag e_{j+1\sigma}^\dag \hat{g}_{j\sigma} 
- \hat{b}_j^\dag e_{j-1\sigma}^\dag \hat{g}_{j\sigma} + H.c. \big) ~.
\end{eqnarray}
We make the mean-field approximation to the boson operator
\begin{equation}
\hat{b}_j=\langle \hat{b}_j \rangle =b_j  \approx b ~,\qquad
\hat{b}_j^\dag=\langle \hat{b}_j ^\dag \rangle =b_j^*  \approx b^* ~.
\end{equation}
The effective action is written as
$S=\int \dd \tau \,\mathcal{L} \equiv \int \dd \tau \,(\mathcal{L}_0+\mathcal{L}_1+\mathcal{L}_2)$ with
\begin{eqnarray}
& \mathcal{L}_0 = \sum_{j\sigma} \big( 
\hat{g}_{j\sigma}^\dag \partial_\tau \hat{g}_{j\sigma} 
+ e_{j\sigma}^\dag \partial_\tau e_{j\sigma} \big)
-\tilde{\mathcal{H}} ~,\\
& \mathcal{L}_1 = - \sum_j \lambda_j
\Big( |b_j|^2 + \sum_\sigma \hat{g}_{j\sigma}^\dag \hat{g}_{j\sigma} - 1 \Big) ~,\quad
\mathcal{L}_2=-\Delta_c |\alpha|^2 ~.
\end{eqnarray}
Here $\mathcal{L}_1$ is introduced by the constraint from the particle number conservation (\ref{eq-app-constraint})
with the Lagrange multiplier $\lambda_j$,
and $\mathcal{L}_2$ is induced from photons.
The effective Hamiltonian is represented by combining $\tilde{\mathcal{H}}$ and $\mathcal{L}_1+\mathcal{L}_2$
\begin{eqnarray}
\mathcal{H}_\mathrm{eff} =&~ E_0+t_g\sum_{\langle ij \rangle,\sigma} \hat{g}_{i\sigma}^\dag \hat{g}_{j\sigma}
-t_e\sum_{\langle ij \rangle,\sigma} e_{i\sigma}^\dag e_{j\sigma}
+\sum_{j,\sigma}(\lambda_j-\Delta_{a})\hat{g}_{j\sigma}^\dag \hat{g}_{j\sigma} \nonumber\\
&+ \alpha K\sum_{j,\sigma}\big( b_j^* e_{j+1\sigma}^\dag \hat{g}_{j\sigma} 
- b_j^* e_{j-1\sigma}^\dag \hat{g}_{j\sigma} + H.c. \big) ~.
\end{eqnarray}
where $E_0\equiv \sum_{j} \lambda_j ( |b_{j}|^2 -1)+\Delta_c |\alpha|^2$.
The Lagrange action is then written as $\mathcal{L}= \sum_{j\sigma} \big( \hat{g}_{j\sigma}^\dag \partial_\tau \hat{g}_{j\sigma} 
+ e_{j\sigma}^\dag \partial_\tau e_{j\sigma} \big) -\mathcal{H}_\mathrm{eff}$.
$\mathcal{H}_\mathrm{eff}$ can be diagonalized into a quadratic form
\begin{equation}
\mathcal{H}_\mathrm{eff}=\sum_{\eta=1}^{4L} E_\eta \alpha_\eta^\dag \alpha_\eta + E_0
\end{equation}
by employing the Bogoliubov-de Gennes (BdG) transformation
\begin{equation}
\hat{g}_j=\sum_{j=1}^{2L} u_j^\eta \alpha_\eta ~,\qquad
e_j=\sum_{j=1}^{2L} v_j^\eta \alpha_\eta ~.
\end{equation}
Here $E_\eta$ is the BdG quasiparticle spectrum.
The coefficients $u^\eta\equiv(u^\eta_1,\cdots,u^\eta_{2L})^T$
and $v^\eta\equiv(v^\eta_1,\cdots,v^\eta_{2L})^T$
obey the BdG equation expressed as
\begin{equation}
\mathcal{H}_\mathrm{BdG}
\left(\begin{array}{c}
u^\eta \\
v^\eta
\end{array}\right)
=E_\eta
\left(\begin{array}{c}
u^\eta \\
v^\eta
\end{array}\right) ~,\quad
\mathcal{H}_\mathrm{BdG}=\left(\begin{array}{cccc}
\hat{h}_{g\uparrow} & 0 & \hat{v}^\dag & 0 \\
0 & \hat{h}_{g\downarrow} & 0 & \hat{v}^\dag \\
\hat{v} & 0 & \hat{h}_{e\uparrow} & 0 \\
0 & \hat{v} & 0 & \hat{h}_{e\downarrow}
\end{array}\right) ~,
\end{equation}
where $[\hat{h}_{g\sigma}]_{ij} = t_g \delta_{i,j\pm1} + (\lambda_j-\Delta_a)\delta_{ij}$,
$[\hat{h}_{e\sigma}]_{ij} = -t_e\delta_{i,j\pm1}$,
and $\hat{v}_{ij} = \pm \alpha K b_j^* \delta_{i,j\pm1}$.
The mean-field variables $b_j\approx b$ and $\lambda_i\approx\lambda$ are determined by minimizing the free energy \cite{kondo-book}
\begin{equation}
F=E_0-\frac{1}{\beta}\sum_{k,\alpha}\ln \big[1+\ee^{-\beta E_{\alpha}(k)}\big] ~,
\end{equation}
where $E_{\alpha}(k)$ are eigenvalues of $\mathcal{H}_\mathrm{BdG}$ in the momentum space, and $\beta\equiv 1/T$.
This is achieved by self-consistently solving the following equations
\begin{equation}
\frac{ \partial F}{\partial \theta} =0 \qquad ( \theta=b,\lambda )
\end{equation}
as well as the steady-state condition of the cavity (see Eq.(\ref{eq-steady})) which in the tight-binding model is formulated as
\begin{equation}
\alpha=\frac{\eta}{\Delta_c+\ii\kappa} ~,\qquad
\eta = \frac{K}{L}\sum_{j,\sigma} b_j\big( \langle \hat{g}_{j+1,\sigma}^\dag e_{j\sigma}\rangle
-\langle \hat{g}_{j-1,\sigma}^\dag e_{j\sigma}\rangle \big)~.
\end{equation}

\end{document}